\documentclass[allclo]{FBSart}
\usepackage{amsfonts}
\usepackage{amsmath}
\usepackage{amssymb}
\usepackage{multirow}
\usepackage{bm,graphicx,graphics}
\usepackage{wrapfig}

\title{ The helium trimer with soft-core potentials}

\author{A. Kievsky\instnr{1}, E. Garrido\instnr{2}, C. Romero-Redondo\instnr{2}, and P. Barletta\instnr{3}}
\instlist{
 Istituto Nazionale di Fisica Nucleare, Largo B. Pontecorvo 3, 56127 Pisa, Italy
	\and
Instituto de Estructura de la Materia, CSIC, Serrano 123, E-28006 Madrid, Spain
	\and
Department of Physics and Astronomy, University College London,
Gower Street, London WC1E 6BT, United Kingdom
}

\runningauthor{A. Kievsky}
\runningtitle{The helium trimer with soft-core potentials}

\newcommand {\be}{\begin{equation}}
\newcommand {\ee}{\end{equation}}
\newcommand {\bea}{\begin{eqnarray}}
\newcommand {\ea}{\end{eqnarray*}}
\newcommand {\ba}{\begin{eqnarray*}}
\newcommand {\eea}{\end{eqnarray}}

\newcommand {\bra}{\langle}
\newcommand {\ket}{\rangle}

\newcommand {\ab} {{\it{ab initio}}}
\newcommand {\au} {a.u.}

\begin{document}

\maketitle

\begin{abstract}
The helium trimer is studied using two- and three-body soft-core
potentials. Realistic helium-helium potentials present an
extremely strong short-range repulsion and support a single,
very shallow, bound state. The description of systems 
with more than two helium atoms is difficult due to the very large 
cancellation between kinetic and potential energy. We analyze the possibility
of describing the three helium system in the ultracold regime using
a gaussian representation of a widely used realistic potential, the LM2M2
interaction. However, in order to describe correctly the trimer ground state
a three-body force has to be added to the gaussian interaction. With this
potential model the two bound states of the trimer and the low energy 
scattering helium-dimer phase shifts obtained with the LM2M2 potential
are well reproduced.
\end{abstract}

\maketitle

\section{Introduction}

In recent years systems of two- and three-helium atoms have been object of
intense investigation from a theoretical and experimental
point of view. The existence of the He-He molecule was
experimentally established in the nineties using diffraction 
experiments~\cite{luo2,schol,schol2,gris}. Its binding energy has been
estimated to be around 1 mK and its scattering length $a_0$ around
190 a.u. This makes the He-He molecule one of the biggest diatomic
molecules. Parallel to these studies, several accurate investigations of the 
He-He interaction have appeared in the literature. We can
mention the potentials called 
HFDB \cite{aziz3}, LM2M2 \cite{aziz1}, TTY \cite{tang1}
and the potentials SAPT1 and SAPT2 \cite{aziz4}, 
constructed on a completely \ab \ calculation made by Korona
et al.~\cite{kor1}, using infinite order perturbation theory (SAPT) and a very
large orbital basis set. In addition, retarded dipole-dipole dispersion
interaction is included over the range $0-100000$ \au (SAPT1), or the more
appropriate 10-100000 \au (SAPT2). Recently, in Refs.~\cite{jez07,cen07},
He-He interactions including retardation corrections and a
non-additive three-body term have also been determined. All these
potentials present the common feature of a sharp repulsion
below an inter-particle distance of approximately 5 \au.

Another important characteristic of the He-He interactions is their
effective range $r_0$, which takes a value of around $13$ \au. Accordingly, the
ratio $a_0/r_0$ is rather large ($>10$), which has important consequences in the 
properties of the three-atom systems. In fact, the (bosonic) three $^4$He system presents an
excited state just below the atom-dimer threshold that has been identified
as an Efimov state~\cite{ef2,ef1}. As shown by Efimov, when at least two of the
two-body subsystems present an infinitely large scattering length (or zero
binding energy) an infinite sequence of bound states (called Efimov states)
appear in the three-body system. The number of these states has been
estimated in Ref.~\cite{ef1} to be $N=(\omega_0/\pi){\rm ln}|a_0/r_0|$, with
$\omega_0\approx 1.00662378$ (see Ref.~\cite{kol09} and references therein).
Triggered by this interesting fact, several investigations of the helium
trimer have been produced establishing that its excited state is indeed an
Efimov-like state (see for example Refs.~\cite{esry,bar01,nie98}). In addition,
analysis of the atom-dimer collision in the ultracold regime have
also been performed ~\cite{kol09,mot01,sun08}.

Specific algorithms have been developed to solve the quantum
mechanical three-body problem. In the particular case of three helium atoms,
due to the strong repulsion of the He-He potential, the Faddeev equation has been
opportunely modified~\cite{kol98}. Also, the hyperspherical adiabatic (HA)
expansion has been extensively used in the description of three-body
systems (for a review see Ref.~\cite{nie01}). However, due to the
difficulties in treating the strong repulsion, few calculations
exist for systems with more than three helium atoms. For example,
in Ref.~\cite{lew} the diffusion Monte Carlo method was used to
describe the ground state of He molecules up to 10 atoms. On the
other hand, description of few-atoms systems using soft-core
potentials are currently operated (see for example Ref.~\cite{ste09}).
Therefore the question of the equivalence between a description
using a hard-core potential or a soft-core one needs some
clarification. 

Accordingly, in the present work we shall discuss the description of the 
three $^4$He system using an
attractive He-He gaussian potential designed to reproduce
the helium dimer binding energy, the He-He scattering
length $a_0$, and the effective range $r_0$ of the LM2M2 potential \cite{aziz1}.
Two potentials having similar values of $a_0$ and $r_0$ predict
similar phase shifts in the low energy limit and, therefore, even
if their shape is completely different, they describe
in an equivalent way the physical processes in that limit. It is 
clear that when the energy of the system is increased the details of the
potential become more and more important. In a three-body system
the relative energy of a two-body subsystem is not fixed and,
depending on the particular structure under observation, a certain
range of those two-body energies could be of importance in the
construction of such a structure. When this range exceeds the region of 
equivalence of the two potentials, the three-body structures obtained
with them could be different.
A clear example is the ground state energy of the three-helium system,
which takes the value of $-126.4$ mK when the LM2M2 potential is used, 
and $-150.0$ mK when using its gaussian representation.
It is clear that the lack of repulsion in the gaussian potential
allows the atoms to be closer increasing their binding energy. 

In order to make the description with the gaussian potential closer
to the one obtained when using the LM2M2 potential, we shall include 
a three-body force of short-range character, such that its strength 
will be fixed to reproduce the ground state binding energy of the
system. The quality of this description will be judged by
comparing the binding energy of the 
excited Efimov state and the low energy helium-dimer phase shifts to those 
obtained with the LM2M2 potential. As we will see, both descriptions are
in extremely good agreement showing that it is possible to replace the
potential presenting a sharp repulsion by opportunely designed
two- and three-body soft-core potentials to explore the low-energy regime. 

The numerical calculations will be performed by means of the hyperspherical
adiabatic expansion method as a tool to obtain three-body wave functions \cite{nie01},
and the recently introduced integral relations in order to extract the phase shifts
in 1+2 reactions \cite{bar09,kie10}. A brief summary of these two methods will
be given in the next section. In section \ref{sec3} we give the details of the soft-core
gaussian He-He interaction, and discuss the results obtained for the 
three-$^4$He system when such interaction is used. As we shall show, the use of the integral
relations will permit to extract accurate phase shifts without requiring a correct
description of the asymptotic part of the three-body wave function. 
The role played by the inclusion of the short-range three-body force is investigated in 
section~\ref{sec4}. We close the paper with the summary and the conclusions.

\section{Scattering states using the HA expansion}

\subsection{Phase shifts and integral relations}

Scattering states are usually investigated by determining the asymptotic part of the 
scattering wave function, from which the phase shifts (or the ${\cal K}$-matrix) are extracted.
However, even for processes involving only three particles, the calculation of the large distance
part of the wave function is a delicate issue. Very often a proper description of this 
asymptotic part requires an extremely large basis set,  
which makes the problem sometimes unaffordable.

In Refs.\cite{bar09,kie10} a new method intended to extract phase shifts for 1+$N$ reactions was 
presented. This method, which is derived from the Kohn Variational Principle, is a generalization 
to more than two particles of the integral relations shown in \cite{har67,hol72}, and it permits to 
extract phase shifts from the internal part of the wave functions. In other words, knowledge of the 
large distance asymptotics is not needed, and therefore accurate calculations can be performed 
by use of a basis of a much smaller size. Here we summarize the method for 1+2 reactions where 
only the elastic channel is open, although it can be generalized to multichannel processes \cite{rom10}.

Let us consider a process where a particle hits a bound dimer (1+2 reaction), and let us assume
that the incident energy is  below the threshold for breakup of the two-body bound target. 
This is a three-body process that can be described
through the usual Jacobi coordinates $\bm{x}$ and $\bm{y}$ \cite {nie01}, where for convenience the 
$\bm{x}$-coordinate is chosen between the two particles forming the dimer. 
The three-body wave function $\Psi$ describing
the corresponding three-body system is given by the solution of the Schr\"{o}dinger equation:
\begin{equation}
\left( {\cal H}-E \right) \Psi=
\left( -\frac{\hbar^2}{2m}\bm{\nabla}^2+V-E \right) \Psi=0,
\label{Schrod}
\end{equation}
where $m$ is the normalization mass used to describe the Jacobi coordinates \cite{nie01}, 
$V$ represents the sum of the interactions between each pair of particles, and $E$ is the three-body energy.

For our particular case, where the energy of projectile is below the threshold for breakup of the
dimer, the asymptotics of the three-body wave function takes the form:
\begin{equation}
\Psi \rightarrow AF+BG=A\left( F+\frac{B}{A}G\right),
\label{asym}
\end{equation}
where
\begin{eqnarray}
F&\!\!\!\!=\!\!\!\!&\sqrt{k_y}j_{\ell_y}(k_y y)
\left[ \psi^{j_x} \otimes
\left[Y_{\ell_y}(\Omega_y) \otimes \chi_{s_y} \right]^{j_y}\right]^{JM}
\nonumber
\\
G&\!\!\!\!=\!\!\!\!&\sqrt{k_y}\eta_{\ell_y}(k_y y)
\left[ \psi^{j_x} \otimes
\left[Y_{\ell_y}(\Omega_y) \otimes \chi_{s_y} \right]^{j_y}\right]^{JM},
\label{fg}
\end{eqnarray}
where $j_{\ell_y}$ and $\eta_{\ell_y}$ are the regular and irregular spherical Bessel
functions, $\ell_y$ is the relative angular momentum between projectile and target, 
$\psi^{j_x}(\bm{x})$ is the wave function of the bound dimer with angular momentum $j_x$,
and $\chi_{s_y}$ is the spin function of the projectile. The quantum numbers $\ell_y$ and
$s_y$ couple to $j_y$, which couples to $j_x$ to the total angular momentum $J$ of the three-body
system with projection $M$. Finally, the momentum $k_y$ is given by $\sqrt{2m(E-E_{2b})/\hbar^2}$,
where $E_{2b}$ is the binding energy of the dimer.

From Eq.(\ref{asym})  we can immediately identify:
\begin{equation}
\tan \delta = -\frac{B}{A},
\label{relint}
\end{equation}
and the coefficients $A$ and $B$ are given by \cite{bar09,kie10}
\begin{eqnarray}
        B & = & -\frac{2m}{\hbar^2} 
\left[
  \langle F|{\cal H}-E |\Psi \rangle - \langle\Psi | {\cal H}-E |F 
\rangle \right]  \label{bcoef}\\
        A & = & -\frac{2m}{\hbar^2} 
\left[
  \langle \Psi|{\cal H}-E |G \rangle - \langle G | {\cal H}-E |\Psi 
\rangle
\right], 
\label{acoef}
\end{eqnarray}
where we have made used of the normalization condition:
\begin{equation}
-\frac{2m}{\hbar^2}
\left[
  \langle F|{\cal H}-E |G \rangle - \langle G |{\cal H}-E |F \rangle
\right]=1.
\label{norma}
\end{equation}

When $\Psi$ is the exact solution of the Schr\"{o}dinger equation (\ref{Schrod}), it is then
obvious that $A$ and $B$ are given by:
\begin{eqnarray}
        B & = &  \frac{2m}{\hbar^2} 
  \langle\Psi | {\cal H}-E |F \rangle  \label{intb}\\
        A & = & -\frac{2m}{\hbar^2} 
  \langle \Psi|{\cal H}-E |G \rangle.
\label{inta}
\end{eqnarray}

In Refs.\cite{bar09,kie10,rom10} it is proved that, thanks to the Kohn Variational Principle, 
when a trial three-body wave function $\Psi^t$ is used, the two expressions above are still 
valid up to second order in $\delta \Psi=\Psi-\Psi^t$. This fact permits to extract a second
order approximation of the phase shifts according to Eq.(\ref{relint}). 

It is important to have in mind that the Bessel function $\eta_\ell{_y}$ contained in $G$,
Eq.(\ref{fg}), is not regular at the origin. This technical problem is solved by replacing in the
expressions above the function $G$ by $G(1-e^{-\gamma y})^{\ell_{y}}$, where $\gamma$ is a non-linear 
parameter. The results are stable with $\gamma$ in a range of values around $1/r_0$ where $r_0$ is
the range of the interactions.

As already mentioned, a crucial point is that 
the integral relations in Eqs.(\ref{intb}) and (\ref{inta}) depend only on the
short-range structure of the scattering wave function $\Psi^t$. This is because $F$ and 
$G$ are asymptotically solutions of $({\cal H}-E)F,G=0$. As a consequence, it is not
necessary to compute the trial three-body wave function $\Psi^t$ at very large distances, and
the numerical problem is enormously simplified.

In a general multichannel process the coefficients $A$ and $B$ are actually $n_0 \times n_0$
matrices, with $n_0$ being the number of open channels, and the 
corresponding ${\cal K}$-matrix is given by $A^{-1}B$ \cite{rom10}.

\subsection{The hyperspherical adiabatic expansion method}

When describing 1+2 reactions, a particularly convenient choice for the three-body trial
wave function $\Psi^t$ is the one obtained by use of the adiabatic expansion method.
In this method the wave functions are described by means of the hyperspherical coordinates,
which contain a radial coordinate, the hyperradius $\rho$, and five hyperangles 
$\{\alpha,\Omega_x,\Omega_y\}$. The hyperradius is defined from the Jacobi coordinates
as $\rho^2=x^2+y^2$, while the hyperangle $\alpha$ is given by $\tan \alpha=x/y$, and
$\Omega_x$ and $\Omega_y$ describe the directions of $\bm{x}$ and $\bm{y}$, respectively.

The key of the method is that the hyperangular coordinates vary much faster than $\rho$, in such 
a way that it is possible to solve first the angular part of the Schr\"{o}dinger 
(or Faddeev \cite{nie01}) equation for a set of {\it frozen} values of $\rho$. This amounts
to solve the eigenvalue problem:
\begin{equation}
\left[ \hat{G}^2+\frac{2m\rho^2}{\hbar^2}V(\rho,\Omega)\right] \Phi_n(\rho,\Omega)=
\lambda_n(\rho) \Phi_n(\rho,\Omega),
\label{angu}
\end{equation}
where $\hat{G}$ is the grand-angular operator whose eigenfunctions are the hyperspherical harmonics.

The three-body wave function is then expanded in terms of the basis formed by the complete set
of angular functions $\{\Phi_n(\rho,\Omega)\}$, in such a way that:
\begin{equation}
\Psi(\bm{x},\bm{y}) = \frac{1}{\rho^{5/2}}\sum_{n=1}^\infty f_n(\rho) \Phi_n(\rho,\Omega),
\label{3bwf}
\end{equation}
and the radial functions $f_n(\rho)$ are obtained after solving the coupled set of radial equations \cite{nie01}:
\begin{eqnarray}
\left(
-\frac{\partial^2}{\partial \rho^2}+\frac{\lambda_n(\rho)+\frac{15}{4}}{\rho^2}-Q_{nn}
-\frac{2m(E-W_{3b}(\rho))}{\hbar^2}
\right)f_n(\rho)& = & \nonumber \\
& & \hspace*{-5cm}
= \sum_{n\neq n'} \left(2P_{nn'}\frac{\partial}{\partial \rho}+Q_{nn'}\right) f_{n'}(\rho),
\label{radial}
\end{eqnarray}
where the coupling terms $P_{nn'}$ and $Q_{nn'}$ are given by:
\begin{eqnarray}
& & P_{n n'}(\rho)=\Big\bra \Phi_n(\rho,\Omega) \Big|\frac{\partial}{\partial \rho} \Big|
                           \Phi_{n^\prime}(\rho,\Omega) \Big\ket_\Omega  \nonumber \\
& & Q_{n n'}(\rho)=\Big\bra \Phi_n(\rho,\Omega) \Big|\frac{\partial^2}{\partial \rho^2} \Big|
                           \Phi_{n^\prime}(\rho,\Omega) \Big\ket_\Omega.
\label{coup}
\end{eqnarray}

It is important to note that the coupled equations (\ref{radial}) are actually a set of radial Schr\"{o}dinger
like equations with effective radial potentials:
\begin{equation}
V^{(n)}_{eff}(\rho)=\frac{\hbar^2}{2m}\left( \frac{\lambda_n(\rho)+\frac{15}{4}}{\rho^2}-Q_{nn}(\rho) \right),
\label{effect}
\end{equation}
which contain the eigenvalues $\lambda_n(\rho)$ of the angular part (\ref{angu}). 

The great advantage of using
the hyperspherical adiabatic expansion method is that each adiabatic term in the expansion (\ref{3bwf}) is
associated to a very specific asymptotic structure. In particular, if the three-body system contains one or
more bound two-body subsystems, we have that for each of them one of the angular eigenvalues goes 
asymptotically as
$\lambda_n(\rho) \rightarrow 2m E_{2b}\rho^2/\hbar^2$ \cite{nie01}, which means that its effective 
potential partner (\ref{effect}) goes asymptotically to the binding energy $E_{2b}$ ($<0$) of the corresponding
bound two-body subsystem. Furthermore, it can be proved, see for instance  \cite{nie01}, that 
the angular eigenfunction $\Phi_n(\rho,\Omega)$ associated to the  eigenvalue $\lambda_n(\rho)$ 
corresponds asymptotically to a bound two-body state (the one with binding energy $E_{2b}$), and the 
third particle in the continuum. 

Summarizing, when using the hyperspherical adiabatic expansion method, all the possible elastic, inelastic, 
or rearrangement channels in a 1+2 reaction (or in general in a 1+$N$ reaction) are easily identified.
In fact, they are associated to some specific adiabatic terms. This means that for each incoming
channel, only a reduced amount of adiabatic terms in the expansion (\ref{3bwf}) behave asymptotically 
as (\ref{asym}) and (\ref{fg}). All the others vanish asymptotically, and therefore the size of the 
${\cal K}$-matrix describing the full process is also small. Only breakup processes are described 
by infinitely many adiabatic terms, but this situation will not be considered in this work.

Another important point concerning three-body calculations is that when only pairwise interactions 
are included, the binding energies obtained for bound three-body states typically do not match with
the experimental values. The case of the halo nuclei $^6$He or $^{11}$Li are well known examples \cite{zhu93}. 
The reason for this behavior lies in the fact that three-body correlations, particle polarizations, and in 
general all those effects that go beyond pure two-body correlations, are not taken into account. To solve this
problem, the usual way is to fine tune the binding energies by including an effective hypercentral three-body 
potential, which has been denoted by $W_{3b}(\rho)$ in the radial Eqs.(\ref{radial}). Since this potential
is intended to account for the effects beyond two-body correlations, it should play a role only when
all the three particles are close of each other. This implies that the potential has to be of 
short-range character.

Application of the hyperspherical adiabatic expansion method has proved to be very efficient for the
description of bound states. The convergence of the expansion in Eq.(\ref{3bwf}) is rather fast, and
usually no more than about seven or eight terms are more than enough to get an accurate wave function.
However, when describing scattering states, the convergence of the phase shifts extracted from the
asymptotic part of the wave function slows down dramatically \cite{bar09b}. Even more, extrapolation 
of the computed phase shifts in terms of the number of adiabatic channels included in the calculation could 
lead to a quite inaccurate value. Nevertheless, as shown in \cite{bar09,kie10}, this problem disappears
when the phase shift is obtained through the integral relations (\ref{intb}) and (\ref{inta}). In this
case, since they are obtained from the internal part of the wave function, the convergence
is as fast as for bound states. Thanks to this, the hyperspherical adiabatic expansion method appears
as a very efficient way of computing the trial three-body wave function to be used in the integral
relations. The convergence of the computed phase shift is fast, the size of the basis required in the
calculations remains within affordable limits, and the clean distinction between the different
open channels provided by the adiabatic approximation can then be exploited.

\section{The helium trimer using a two-body gaussian potential}
\label{sec3}

\subsection{The He-He interaction}

As mentioned in the introduction, in this work we shall describe the $^4$He-$^4$He interaction
 by means of gaussian soft-core potential. Following \cite{nie98}, its form is chosen as 
\begin{equation}
V_{2b}(r)=-1.227 \; e^{-r^2/10.03^2} , 
\label{pot2b}
\end{equation}
where $r$ is given in a.u. and the strength is in K. This potential leads to a bound 0$^+$
$^4\mbox{He}_2$ dimer with binding energy $E_{2b}=-1.2959$ mK, scattering length $a_0=189.947$ a.u.,
and effective range $r_0=13.846$ a.u..
This potential was built to provide a good agreement with the more realistic and 
sophisticated hard-core LM2M2 potential \cite{aziz1}, in particular for the binding energy of the
 Helium dimer and He-He scattering at low energy.  
The corresponding LM2M2 values are $E_{2b}=-1.302$ mK, $a_0=189.054$ a.u. and $r_0=13.843$ a.u..

In order to investigate the energy range where the equivalence of the two potentials holds, 
 we define the effective range function as
\begin{equation}
K(E)= k\cot\delta \; ,
\label{efrang}
\end{equation}
where $k^2= M_{He} E/\hbar^2$ ($M_{He}$ is the mass of the He-atom and $E$ is the two-body center of
mass energy),  and $\delta$ is the $s$-wave 
helium-helium phase shift. For low energy values this function is known to take the form:
\begin{equation}
K(E\rightarrow 0)\longrightarrow -\frac{1}{a_0} + \frac{1}{2}r_0 k^2 \;\; .
\label{lowen}
\end{equation}

\begin{figure}[t!]
\begin{center}
\includegraphics[width=10cm,angle=0]{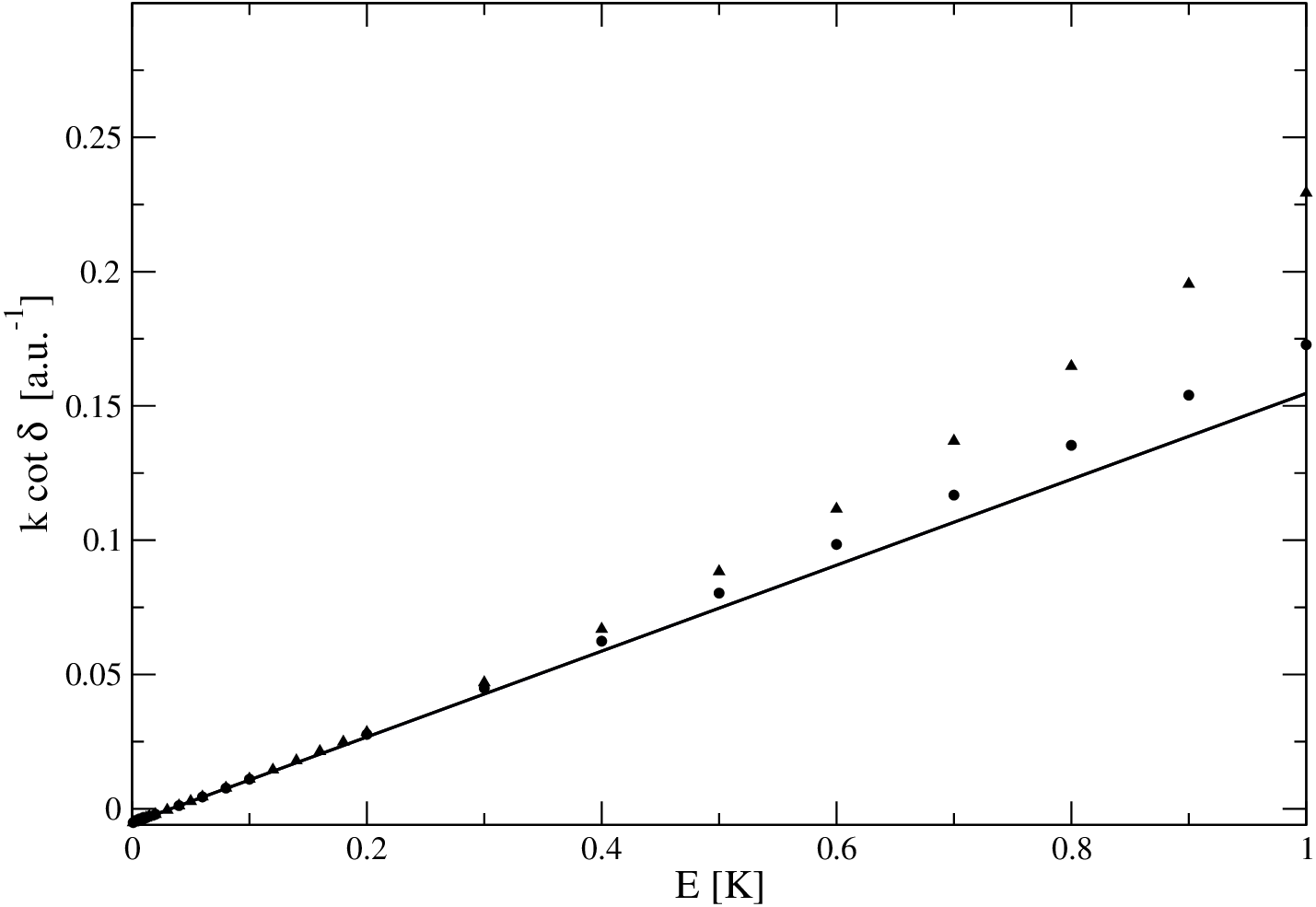}
\end{center}
\caption{The effective range function of the LM2M2 potential (triangles)
and its gaussian representation (circles) as a function of the two-body
center of mass energy. The low energy limit of the effective range function 
is given as a solid line.}
\label{fig:fig1}
\end{figure}

In Fig.~\ref{fig:fig1} the phase shifts of the LM2M2 (triangles) and
gaussian (circles) potentials are given in the form of the effective range function
(\ref{efrang}) as a function of the two-body energy $E$.
As a reference the straight line of the low energy representation of
$K(E)$, Eq.(\ref{lowen}), is also displayed. From the figure we can see that the phase shifts
of both potentials follow the low energy limit of $K(E)$ up to $0.2$ K 
approximately. Above this value the phase shifts do not follow 
the straight line and above $0.4$ K they start to be different for both potentials.

\subsection{The Helium trimer}

In this section we investigate elastic He-He$_2$ collisions at energies below the dimer 
break-up threshold. The energy range of interest is thus $E_{2b}<E<0$, where $E_{2b}$ is the
binding energy of the dimer.
 
 We compute the atom-diatom phase shifts by means of the integral relations
 given in Eqs.(\ref{intb}) and (\ref{inta}), and which permit to obtain the phase shift as given in (\ref{relint}). 
The advantage of using this approach is that the accuracy of the result depends only on
 the internal part of the wave-function, and knowledge of its asymptotic behaviour is not 
necessary. To illustrate this point the continuum wave functions for the 
He-He$_2$ system are obtained by imposing a simple box boundary condition to the coupled set of
hyperradial equations (\ref{radial}), that is $f_n(\rho)=0$ for $\rho \geq \rho_{max}$,
where $\rho_{max}$ determines the size of the box. In other words, we solve the Eqs.(\ref{radial})
by putting an infinite wall in all the effective adiabatic potentials (\ref{effect}) at $\rho_{max}$.

A direct consequence of imposing a box boundary condition is that the continuum spectrum
of the system is discretized. Only three-body energies associated to wave functions
that are zero at the wall of the box are allowed. The energies of the discretized
spectrum change in values when $\rho_{max}$ is changed, and 
increase in number and density as 
$\rho_{max}$ increases. This is contrary to what happens
to a bound state, whose wave function always vanishes at the wall of the box provided that
the box is big enough to hold it. 

In order to study the convergence of the phase shift 
calculation in function of the size of the box, we have chosen four values of $\rho_{max}$, 
namely  420.976 a.u., 1619.436 a.u, 2916.698 a.u, and 4221.912 a.u., which all give rise to 
a discrete continuum state with a three-body energy of $-0.7959$ mK. This corresponds to an incident
energy for the $^4$He projectile of 0.5 mK. In Table~\ref{tab:table0} the negative spectrum in the
energy region $E_{2b}<E<0$ of the different boxes is shown. For ease of reference,
 the boxes are numbered according to 
their increasing size. The two smaller boxes present only one negative 
eigenvalue, whereas the third box has three and the fourth box has four. In the
two larger boxes the selected eigenvalue of $-0.7959$ mK is the second one. 

\begin{table}[h]
\caption{Negative eigenvalues (in mK) above $E_{2b}$ obtained for the four different boxes}
\label{tab:table0}
\begin{center}
\begin{tabular}{@{}ccccc}
\hline
 box & 1 & 2 & 3 & 4 \cr
 $\rho_{max}$ [a.u.] & 420.976 & 1619.436 & 2916.698 & 4221.912 \cr
\hline
 &-0.795891 & -0.795891 & -0.178999 & -0.409160 \cr
 &          &           & -0.795891 & -0.795891 \cr
 &          &           & -1.171742 & -1.074084 \cr
 &          &           &           & -1.240771 \cr
\hline
\end{tabular}
\end{center}
\end{table}

\begin{figure}[htb]
\begin{center}
\begin{tabular}{c}
\includegraphics[width=6cm,angle=0]{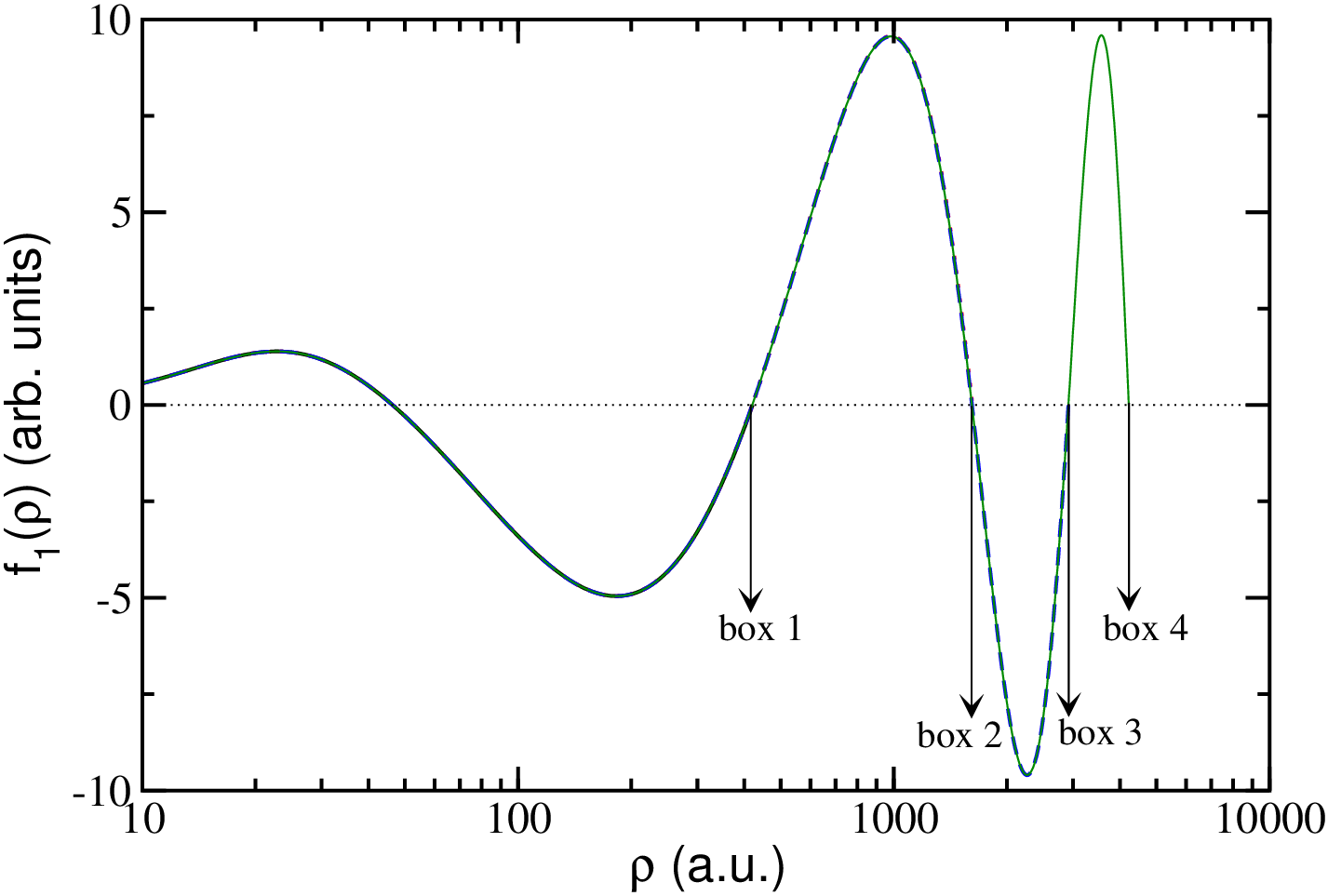} \\
\includegraphics[width=6cm,angle=0]{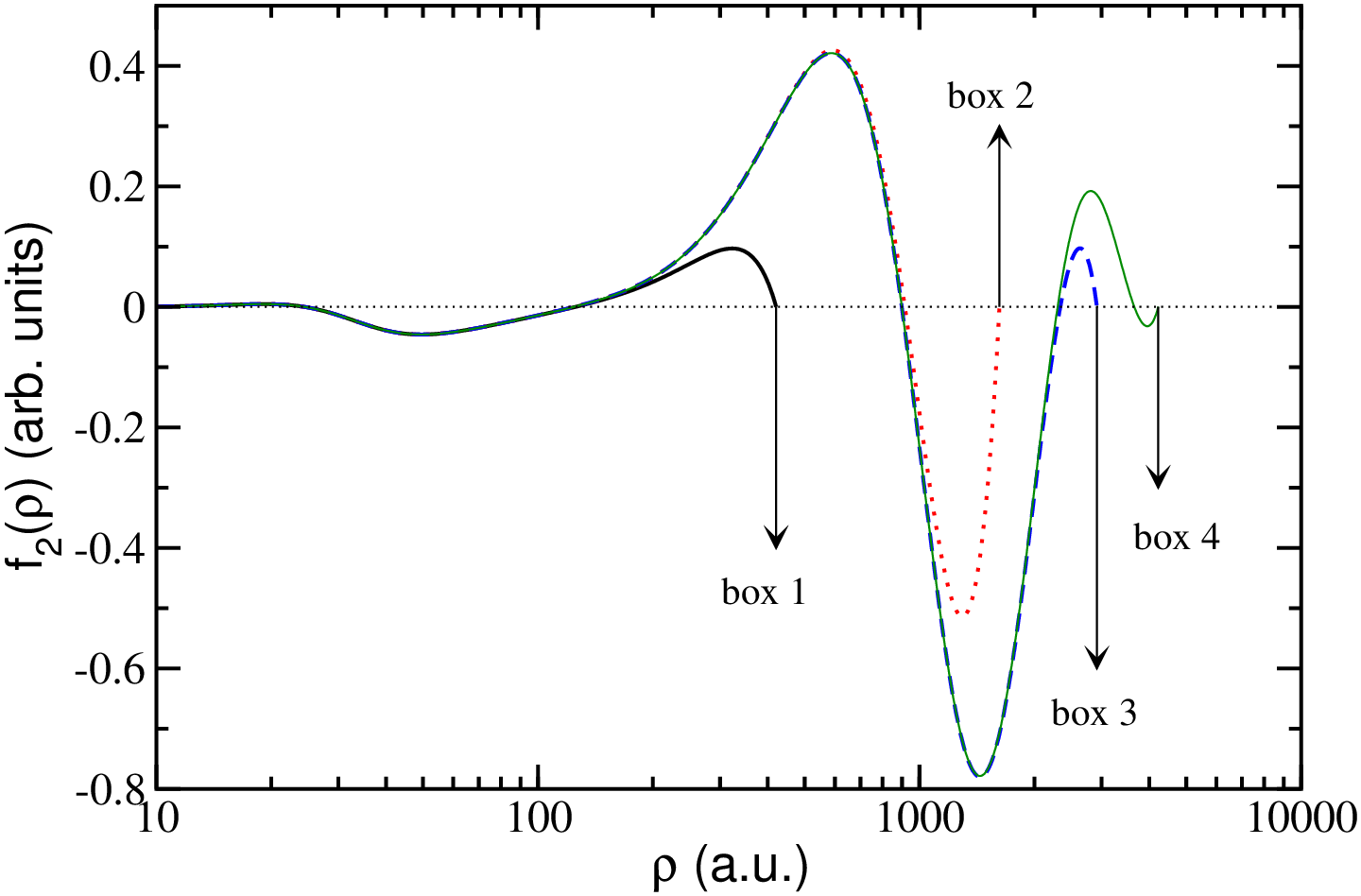} \\
\includegraphics[width=6cm,angle=0]{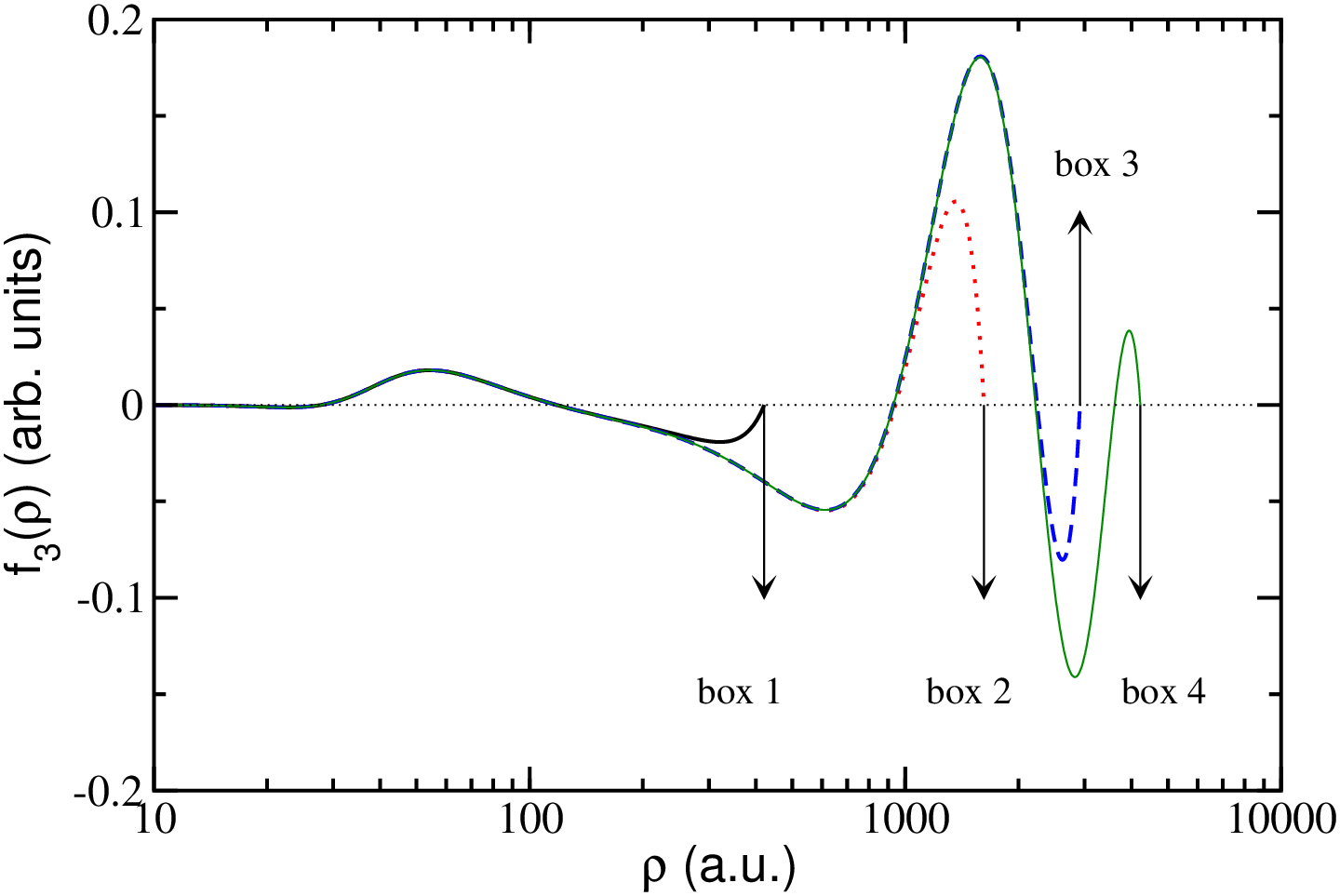} 
\end{tabular}
\end{center}
\caption{(Color online) Three first radial wave functions for the He-He$_2$ system at a three-body energy of 
$-0.7959$ mK, which corresponds to an incident energy of the $^4$He projectile of 0.5 mK. They have
been computed imposing a box boundary condition.
The wave functions have been obtained with four different values of $\rho_{max}$ (see text)
such that all of them give rise to the same selected discretized continuum state. The thick
solid, dotted, dashed, and thin solid curves are the radial wave functions corresponding to the
boxes 1, 2, 3, and 4, respectively.}
\label{fig:box}
\end{figure}

 Fig.\ref{fig:box} shows the first three hyperradial 
wave functions $f_n(\rho)$ ($n=1,2,3$) obtained by solving the system (\ref{radial})
 at energy $E=-0.7959$ mK for each of the four boundary conditions discussed above, namely, 
 ``box 1'' (thick solid curve), ``box 2'' (dotted curve), ``box 3'' (dashed curve),
 and ``box 4'' (thin solid curve). As we can see, all the hyperradial wave functions perfectly
 overlap, except in proximity of their respective $\rho_{max}$ value, where they begin to bend in order to reach zero
all of them at $\rho_{max}$. 

Once the radial wave functions for the ``box 1'', ``box 2'', ``box 3'', and ``box 4'' calculations
have been obtained, it is possible to compute the corresponding atom-diatom phase shift
by using the integral relations. When the calculation is performed,
 we obtain for the four cases a phase shift value of $-47.19$
degrees (``box 1''), $-40.20$ degrees (``box 2''), $-40.54$ degrees (``box 3'') , and $-40.54$ degrees
(``box 4'').  These results have been obtained using 30 adiabatic terms in the 
expansion (\ref{3bwf}), which are more than enough to get a converged result. The same 
result is actually obtained when about half of the adiabatic terms are considered. 
 As we can see, the first two boxes used are too small, since
the three-body wave function is set equal to zero at $\rho$ values where it still contributes  
to the integrands of the relations (\ref{intb}) and (\ref{inta}).
 For a sufficiently large box the phase shift stabilizes at a value
of $-40.54$ degrees. This value agrees with the result of Ref. \cite{rom10}, 
where the same reaction with the same incident energy
and the same gaussian two-body potential is investigated, but with a three-body wave function
showing the correct asymptotic behavior. 
These results confirm and justify our approach. In fact, we notice that as expected, by 
using the integral relations,  an accurate
 phase shift can be extracted by a wave-function which is accurate in the internal region, but 
which is 
completely inaccurate in the external asymptotic region. In fact, in our calculation
 the asymptotic part is simply chosen equal to zero.

\section{Hypercentral three-body force}
\label{sec4}

A significant difference between the gaussian potential (\ref{pot2b}) and the LM2M2 potential 
 appears in the description of the helium trimer bound states.
 Table~\ref{tab:table1} presents the binding energy for the two helium trimer 
bound states, $E_0$ and $E_1$, as well as the atom-diatom scattering length $a_0$.
 The gaussian potential leads to two bound states
 of energies $-150.0$ mK and $-2.467$ mK. The LM2M2 potential also supports two bound states, but with 
energies $-126.4$ mK and $-2.265$ mK. The gaussian potential thus significantly overbounds the 
two states. This phenomenon can be easily understood as, lacking the hard core, it leads 
to more compact structures where the three particles are closer to each other.
 Though smaller, the difference in the binding energy of 
the shallow excited state is also appreciable. It should be noticed
that the structure of this state corresponds to a two-atom bound structure with the third 
atom orbiting far away. If the attraction of the two-body potential is increased
the two-atom bound structure reduces its size and the third atom
evaporates. This is the mechanism from which the Efimov-like states
disappear from the spectrum when the two-body potential results more
attractive. 
 Finally, there is also a noticeable difference in the atom-diatom 
scattering length $a_0$ calculated with the LM2M2 potential
(from Ref.~\cite{kol09}) and with the gaussian potential.

\begin{table}[h]
\caption{The ground state $E_0$, the excited state $E_1$, and the
helium-dimer scattering length $a_0$ calculated with the LM2M2 potential
and with its gaussian representation. In the last four rows, the results of
the gaussian potential plus the three-body forces are given.}
\label{tab:table1}
\begin{center}
\begin{tabular}{@{}cccc}
\hline
  potential & $E_0$ [mK] &  $E_1$ [mK]  & $a_0$ [a.u.] \cr
\hline
 LM2M2 \cite{kol09}   & $-126.4$ & $-2.265$  & 217.3  \cr
 gaussian & $-150.0$ & $-2.467$  & 165.9 \cr
\hline
 ($W_0$ [K], $\rho_0$ [a.u.]) &       &         &  \cr
 $(306.9,4)$    & $-126.4$ & $-2.283$   & 211.7 \cr
 $(18.314,6)$   & $-126.4$ & $-2.287$   & 210.6 \cr
 $(4.0114,8)$   & $-126.4$ & $-2.289$   & 210.0 \cr
 $(1.4742,10)$  & $-126.4$ & $-2.292$   & 209.2 \cr
\hline
\end{tabular}
\end{center}
\end{table}

Similarly to what done in standard nuclear three-body calculations, where the disagreement
between computed and experimental binding energies is corrected by
the inclusion of an effective three-body force, here we shall investigate
the possibility of correcting the discrepancy between the binding energies
shown in Table~\ref{tab:table1} by adding an analogous three-body
force to the gaussian potential in the description of the helium trimer.
We require that the range of the effective three-body potential $W_{3b}$ be of the
order of the size of the trimer in its ground state. Therefore this
force will help to fix the proper scale in the three-body system given
by the physics (not observed) included in the repulsion of the original
LM2M2 potential. We propose the following simple two-parameter 
hyperradial three-body force
\begin{equation}
W(\rho)=W_0 {\rm e}^{-\rho^2/\rho_0^2} \;\; .
\end{equation}

We analyze four different values for $\rho_0$, namely, $4$, $6$, 
$8$ and $10$ a.u.. For each $\rho_0$, the strength
of the force $W_0$ has been fixed to reproduce the value of
$-126.4$ mK given by the LM2M2 potential for the trimer ground state.
This results in four different pairs $(W_0,\rho_0)$. The
corresponding calculated values for $E_0$, $E_1$ and $a_0$ are given in the last 
rows of Table~\ref{tab:table1}.  The results for $E_1$ and
$a_0$ are predictions and, as we can observe from the table, the inclusion 
of the three body force brings them 
much closer to the values of the LM2M2 potential in all four cases.

\begin{figure}[t!]
\begin{center}
\includegraphics[width=10cm,angle=0]{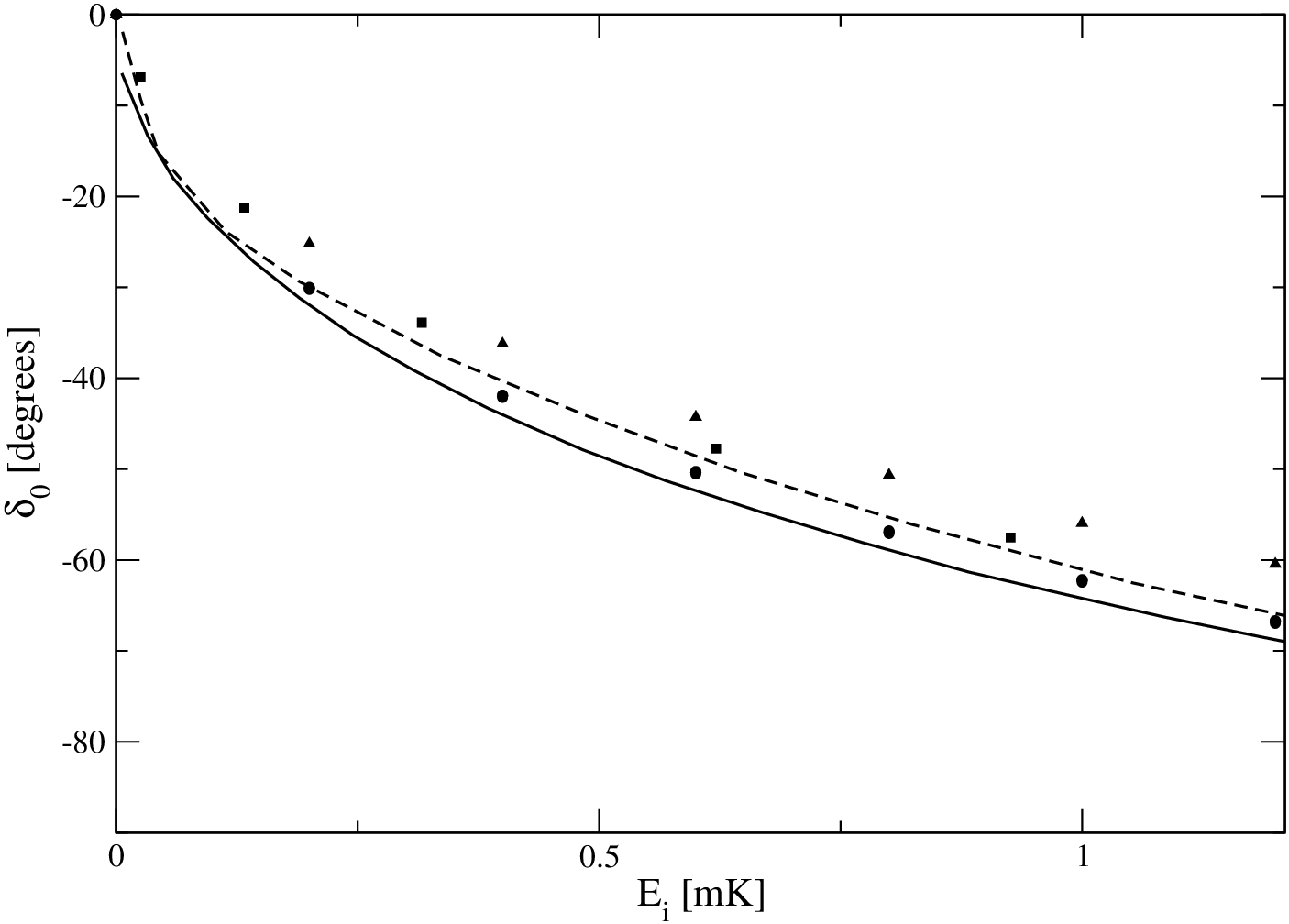}
\end{center}
\caption{Phase shifts for the collision between a $^4$He atom and
a $^4$He$_2$ dimer as a function of the incident center of mass energy. The 
solid curve, solid squares, and dashed curve are the results given in 
\cite{sun08}, \cite{rou03}, and \cite{mot01}, respectively. The solid triangles
are obtained with the He-He gaussian interaction (\ref{pot2b}) without inclusion
of a three-body force. The solid circles show the result with the same gaussian
potential and including the three-body force with the strengths and ranges given in 
Table~\ref{tab:table1}. All the four cases shown in the table give rise to indistinguishable
results.}
\label{fig:fig2}
\end{figure}

The analysis can be extended to the description of atom-diatom 
scattering states below the breakup threshold in three atoms. Using the
technique described in the previous sections, the $s$-wave phase shift
$\delta_0$ has been calculated for different values of the incident center of 
mass energy $E_i=E-E_{2b}$, where $E$ is the total (negative) energy 
of the system and $E_{2b}$ is the dimer energy. The results are shown
in Fig.~\ref{fig:fig2}, where they are compared to the calculations in 
Refs.~\cite{sun08,rou03,mot01}, which are given by the solid line, solid squares
and dashed line, respectively. The results for the gaussian potential
with and without the three-body force are represented the solid circles
 and solid triangles, respectively. 
 In fact, the phase shifts obtained 
using the four parametrizations overlap and they are practically indistinguishable.
As we can see, inclusion of the effective hypercentral three-body force
leads to results very close to the ones in Refs.~\cite{sun08,mot01}.
We can therefore conclude that the gaussian
potential, constructed to reproduce the low energy spectrum of the
two-helium system given by a realistic potential as the LM2M2 selected
in the present analysis, plus a three-body force, constructed to reproduce
the LM2M2 trimer ground state, reproduces the low energy spectrum of the
three-helium system. 

\section{Conclusions}

In this paper we have discussed the possibility of describing the helium trimer system
using a soft-core interaction. The potential, selected of a gaussian type with two parameters,
has been constructed in order to reproduce the low-energy He-He scattering 
as calculated using the LM2M2 potential. We have shown (see Fig.\ref{fig:fig1})
that approximately above $0.4$ K the equivalence breaks down as the details
of the internal part of the interaction become important. When the study is extended 
to the three-body system, similar problems arise, and in fact at a much earlier stage, since already the 
description of the ground three-body state is significantly different between the two potentials.
Using the LM2M2 potential the trimer binding energy is
around $-126$ mK whereas using its gaussian representation an energy of $-150$ mK is obtained.
It is clear that the short range physics embedded in the repulsive part of the LM2M2
potential is missing in the attractive gaussian potential. The question addressed in this
work is if the inclusion of a repulsive short-range three-body force can recover those
aspects of the dynamics not present in the soft-core potential. To this aim a
repulsive gaussian hyperradial three-body force has been parametrized in order
to reproduce the LM2M2 trimer energy. Four different ranges, from 4 a.u. to 10 a.u.,
have been considered taking into account the fact that the repulsive part of the
$^4$He-$^4$He interaction has a range $r_p\approx 5$ a.u. and the hyperradius corresponding to 
a configuration of an equilateral triangle of side $r_p$ is $\sqrt{3}r_p$.
For each value used for the range the corresponding strength has been fixed and, as it can be observed from
Table~\ref{tab:table1}, the LM2M2 trimer energy has been reproduced. Furthermore we
can observe a much better description of the first excited state whose energy
results to be almost constant for the four different parametrizations used.
A difference of about $0.01$ mK is observed between the largest and the 
shorter ranges. Interestingly, also the atom-dimer scattering length is much
better described when the hypercentral potential is included. With the four
parametrizations a value of around 211 a.u. has been obtained, which
is only $3\%$ lower than the LM2M2 value. 

The atom-dimer phase shifts have been studied with the soft-core potential
with and without the inclusion of the three-body term. When this term is
included we have found that the phase shifts, using the four parametrizations,
are almost identical. Moreover they agree very well with the LM2M2 phase shifts
calculated in Refs.~\cite{rou03,mot01} and with the phase shifts given in
Ref.~\cite{sun08} using a different He-He potential. However this is not
the case when only the gaussian two-body potential is used. Therefore we can conclude 
that it is justified to use soft-core potentials to describe three helium atoms
in the ultracold regime using a potential model that includes a short-range
three-body term constructed to fix the ground state energy at the correct level.

As mentioned in the Introduction, it is very difficult to get an accurate description
for systems with
more than three helium atoms using potentials presenting a strong short-range repulsion.
Ref.~\cite{lew} remains an isolate example of an study in which ground state energies 
of He molecules, up to 10 atoms, are computed using the diffusion Monte Carlo algorithm.
No information about excited states are given in the literature.
On the other hand, the possibility of using soft-core potentials in the description of 
these systems will allow the use of methods that at present cannot be applied.
Accordingly this study has to be considered as a first step in this direction.
Studies at higher energies in the three-body system and in systems with
more than three atoms are at present in progress and they will serve to support the
use of soft-core potentials in particular energy regimes.

\section{Acknowledgments}
This work was partly supported by funds provided by DGI of MEC (Spain)
under contract No.  FIS2008-01301. One of us (C.R.R.) acknowledges
support by a predoctoral I3P grant from CSIC and the European
Social Fund.

\end{document}